%% file: main.tex
\documentclass[10pt,a4paper,conference]{IEEEtran}

\usepackage{cite}
\usepackage{amsmath}
\usepackage{amssymb}
\usepackage{adjustbox}
\usepackage{algorithmic}
\usepackage{fixltx2e}
\usepackage{graphicx}
\graphicspath{ {images/} }

\usepackage{kbordermatrix}

\usepackage{url}
\usepackage{algorithmic}
\usepackage{algorithm}
\newlength\myindent
\usepackage{tikz}
\usetikzlibrary{shapes, arrows, decorations, decorations.markings, backgrounds, calc, arrows.meta, shadows, shadows.blur,positioning,spy,shadings}
\pgfdeclarelayer{myback}
\pgfdeclarelayer{myback2}
\pgfsetlayers{background,myback2,myback,main} 

\usepackage{pgfplots}
\pgfplotsset{width=10cm,compat=newest}
\pgfplotsset{compat=newest} 
\pgfplotsset{plot coordinates/math parser=false} 
\newlength\figureheight 
\newlength\figurewidth 
\usepgfplotslibrary{units}
\usepackage{pgfplotstable}

\tikzstyle{vecArrow} = [{line width=0.5pt, double distance=5pt, arrows={-Implies[length=0pt 0.8 0]}}]
\tikzstyle{vecNoArrow} = [anchor=center,decoration={markings,mark=at position
	1 with },
double distance=5pt, shorten >= 3pt,
preaction = {decorate},
postaction = {draw,line width=1.5pt, white,shorten >= 1.5pt}]
\tikzstyle{vecArrowTUMblue} = [anchor=center,decoration={markings,mark=at position
	1 with {\arrow{Straight Barb[length=6pt,TUMblue]}}},
double distance=5pt, shorten >= 3pt,
preaction = {decorate},
postaction = {draw,line width=1.5pt, white,shorten >= 1.5pt}]
\tikzstyle{innerWhite} = [semithick, white,line width=4pt, shorten >= 4.5pt]
\tikzset{
	triangle/.style = { regular polygon, regular polygon sides=3, rotate=-90,thick,scale=1.2},
	triangleWithText/.style = { regular polygon, regular polygon sides=3, shape border rotate = -90, rotate=0,thick,scale=1.2},
	sum/.style = {draw, circle,thick,scale=1.3},
}

\definecolor{mycolor1}{rgb}{0.00000,0.44700,0.74100}%
\definecolor{mycolor2}{rgb}{0.85000,0.32500,0.09800}%
\definecolor{mycolor3}{rgb}{0.92900,0.69400,0.12500}%
\definecolor{mycolor4}{rgb}{0.49400,0.18400,0.55600}

\usepgflibrary{patterns} 
\usepgflibrary[patterns] 
\usetikzlibrary{patterns} 
\usetikzlibrary[patterns] 

\usepackage{textcomp}
\newcommand\copyrighttext{%
	\footnotesize \textcopyright 2018 IEEE.  Personal use of this material is permitted.  Permission from IEEE must be obtained for all other uses, in any current or future media, including reprinting/republishing this material for advertising or promotional purposes, creating new collective works, for resale or redistribution to servers or lists, or reuse of any copyrighted component of this work in other works.
}
\newcommand\copyrightnotice{%
	\begin{tikzpicture}[remember picture,overlay]
	\node[anchor=south,yshift=10pt] at (current page.south) {\fbox{\parbox{\dimexpr\textwidth-\fboxsep-\fboxrule\relax}{\copyrighttext}}};
	\end{tikzpicture}%
}

\newcommand{\journal}[1]{#1} 
\renewcommand{\journal}[1]{} 

\large{\title{Radio Resource Allocation for Reliable Out-of-coverage V2V Communications}}
\IEEEoverridecommandlockouts
\author{
	\IEEEauthorblockN{
			Taylan \c{S}ahin and
			Mate Boban
		}

		\IEEEauthorblockA{
			Huawei Technologies Duesseldorf GmbH, German Research Center, 80992 Munich, Germany\\
			Email: \{taylan.sahin, mate.boban\}@huawei.com
		}
}

\begin{document}
	\maketitle
	\copyrightnotice
	
	\input{Abstract}
	
	\input{Introduction}
	
	\input{System_Model}

	\input{Proposed_Algorithm}

	\input{Results}

	\input{Conclusions}

	\bibliographystyle{IEEEtran}
	\input{main.bbl}

\end{document}

%% file: Abstract.tex
\begin{abstract}

We explore a new approach to radio resource allocation for vehicle-to-vehicle (V2V) communications in case of out-of-coverage areas that are delimited by network infrastructure. 
By collecting and predicting information such as vehicle velocity, density and message traffic, the network infrastructure ensures reliability of the V2V services. We propose reserving required amount of resources for services that cannot be pre-scheduled (e.g., emergency braking, crash notifications, etc.), and scheduling those services that can be pre-scheduled (e.g., platooning). 
We analyze the resource reservation as a function of target reliability under varying vehicle densities and sizes of out-of-coverage area. For pre-scheduled services, we explore how variations in the vehicle velocities and predictions affect successful transmissions. The results indicate that increase in required reliability does not penalize the system prohibitively. On the other hand, speed prediction errors decrease the transmission success rate considerably, thus calling for a more flexible scheduler design
.
\journal{Moreover, we evaluate the impact that system-level parameters, such as vehicle density and size of the out-of-coverage area, have on the reliability.}
\end{abstract}


%% file: Introduction.tex
\section{Introduction}
\label{Introduction}

Vehicle-to-everything (V2X) communications target safer, smarter and more efficient transport systems, as the key enabler of connected vehicles. Vehicles exchange information with each other via vehicle-to-vehicle (V2V) communications, e.g., by sending safety-critical messages containing their position and velocity, among other data. Efficient delivery of these messages could be assured by the assistance of a network infrastructure. A centralized network entity can have a control over the access of vehicles to the radio resources, in order to ensure a reliable V2V communication \cite{poisson}.

On the other hand, availability of the infrastructure is not always guaranteed. 
Vehicles may travel through an area where the connection to the network infrastructure 
is no longer possible. 
In this case, maintaining a reliable V2V communication remains a challenging task.

\subsection{Related Work}


Among the existing wireless technologies, 3GPP standard LTE-A Release 14 provides support for V2X services, referred as LTE vehicular (LTE-V)~\cite{poisson}. V2V communication takes place on the direct link between the vehicles, referred to as sidelink (SL). Two different modes exist: mode 3, where SL resources are scheduled by the cellular infrastructure in a centralized way; and mode 4, in which vehicles autonomously select the SL transmission resources based on a sensing mechanism on the configured resource pools. Mode 4 pools are configured to geographical zones by the network. In case vehicles have no access to the network, they can use a pre-configured set of resources~\cite{3GPP}. 


Radio resource allocation problem targeting V2X has recently gained significant interest. Performance evaluation of LTE-V mode 4 is studied in \cite{Gozalvez}. In \cite{harri}, vehicles make use of the position information transmitted by other vehicles, in order to choose the resources for V2V transmissions without any network supervision. In \cite{vnc}, the resource pools are created in a time-orthogonal manner, with respect to orthogonal road traffic crossing the intersections. Further, vehicles perform sensing-based resource selection inside each pool. Similarly, in \cite{Springer}, an additional resource pool is allocated exclusively for vehicles inside the intersections. Authors further consider a highway scenario, where time-orthogonal resources are allocated for equal sections along the road, spatially alternating on the two directions. At the same time, a separate resource pool orthogonal in frequency is used by the vehicles driving in the fast lanes.


\subsection{Our Contribution}

We focus on the case where the out-of-coverage area is delimited on all sides by infrastructure, e.g., base stations (BSs), as shown in Fig.~\ref{doca}. We are motivated to explore this case for two reasons: 1) it is particularly interesting for early deployment where there will invariably be coverage gaps; and 2) there are obvious situations where even in full deployment coverage gaps will arise (e.g., due to physical obstructions such as tunnels). The goal of our work is to improve the reliability of V2V communications in the delimited out-of-coverage area (DOCA), by using the surrounding BSs to make better resource allocation decisions.

Our approach differs from the state of the art in that resource allocation for the out-of-coverage vehicles is still performed by the network infrastructure, based on the predictions of vehicle locations inside the DOCA, which, along with propagation conditions, determines the interference on a specific resource. We analyze the performance of resource reservation for non-scheduled services as well as scheduling performance of pre-scheduled services. 

The rest of this paper is organized as follows. In Section \ref{Model} we provide our system model and define the problem. Our considerations for resource reservation and scheduling are described in Section \ref{Algorithm}. Section \ref{Results} presents the results of our simulations. Finally, Section \ref{Conclusion} concludes the paper, and discusses the further related work.

%% file: System_Model.tex
\section{System Model}
\label{Model}

\begin{figure}[!t]
	\centering
	\includegraphics[width=\columnwidth, height=3.75cm]{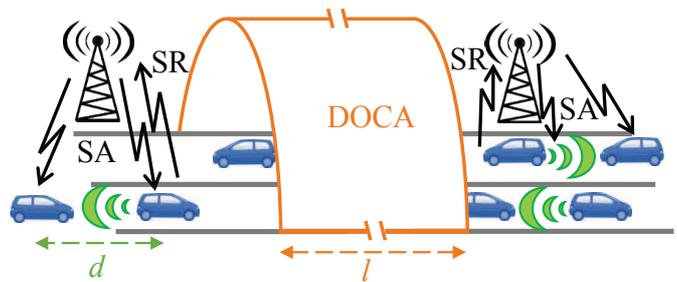}
	\caption{DOCA of length $l$ delimited by BSs on a two-way highway segment, in which vehicles communicate with each other within a distance $d$, by sending SR and receiving SA before they enter it.}
	\label{doca}
\end{figure}

We consider a scenario where vehicles having V2V communications controlled by the BSs, pass through a DOCA during their travels on a two-way highway, with an arrival rate $\lambda_{arr}$ (vehicles per second in two directions). DOCA could be thought of as a tunnel of certain length $l$, where no reception from the BSs is possible, which is illustrated in Fig. \ref{doca}. On the other hand, BSs deployed at each end of DOCA are able to serve the vehicles just before (after) they enter (exit).



We distinguish between two types of V2V services among the vehicles: 1) aperiodic messages transmitted upon a triggering unexpected event for safety-warning purposes, similar to decentralized environmental notification messages (DENMs)~\cite{festag2014cooperative}, which we call ``ad hoc'' services, and 2) messages transmitted with periodicity $T_{p}$ that carry information such as vehicle position and velocity, similar to that of cooperative awareness messages (CAMs)~\cite{festag2014cooperative}, which we see as services that can be ``pre-scheduled''.


We assume that the transmission of each message requires the same amount of radio resources, which we name as a resource block (RB). An RB occupies one specific time slot of length $\Delta t$ which we call transmission time interval (TTI), and one frequency slot of $\Delta f$ called a subchannel, on the assumed radio resource grid, as illustrated in Fig. \ref{sch}.

\subsection{Problem Definition}

We consider that a certain communication reliability is required in the network, e.g., by the V2X services running on the system. A reliable communication between two vehicles is established when the signal-to-interference-plus-noise ratio (SINR) at the receiving (Rx) vehicle is at least equal to or larger than a certain target level. 

For calculating SINR, we employ a model that abstracts the effect of signal propagation. Specifically, we consider SINR as sufficient when the Rx-vehicle is traveling within distance equal or less than distance $d$ away from the transmitting (Tx) vehicle at the time of the transmission, and no other vehicle using the same RB (transmitting at the same subchannel and TTI) is within the same distance $d$ from the Rx-vehicle. We assume that, outside $d$, the power of the transmitted signal becomes too weak compared to the noise power to have a reliable reception, mainly due to propagation losses and fading on the communication channel. On the other hand, if another vehicle is transmitting (receiving) using the same RB at most $d$ away from the Rx- (Tx-) vehicle, the transmissions interfere with each other, making the SINR at the receiver below the desired level. For such reasons $d$ may be also called as ``interference'' or ``broadcast'' range.

Furthermore, it is possible to define the collision domain of our system as the physical space where all transmissions of the vehicles residing inside it would collide or interfere with each other if they were to use the same single RB. In one-dimensional case such as highway, the collision domain is of length $2d$, whereas it occupies an area of $\pi d^2$ in two-dimensional case. For the vehicles, we furthermore impose a half-duplex-communications constraint such that they can not transmit and receive using the same RB.

Our goal is to ensure the reliability of the V2V communications inside the DOCA. Namely, the task is to allocate the radio resources for the transmissions given the constraints of reliability and half-duplex.

%% file: Proposed_Algorithm.tex
\section{Proposed System}
\label{Algorithm}

We propose a centralized entity to manage the radio resources in the network, which particularly requires an access to the road and the message traffic information. The delimiting BSs collect this information from the vehicles entering (exiting) the DOCA. The collected information is then used to make decisions by the centralized controller, e.g., the scheduler. Our solution regarding the radio resource management comprises of two main parts: resource reservation for ``ad hoc'' services in DOCA, and pre-scheduling the services in DOCA. 


\input{analysis}

\subsection{Pre-scheduling the services in DOCA}
\label{Scheduling}

\begin{figure}[!t]
	\centering
	\includegraphics[height=4cm]{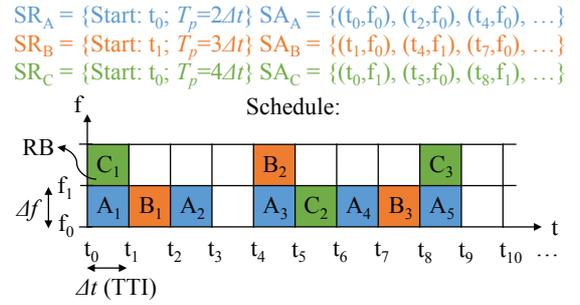}
	\caption{An example schedule on the radio resource grid\journal{(with $F=2$)}, according to SRs sent by 3 vehicles A, B and C. Vehicles in turn, informed by the SAs.}
	\label{sch}
\end{figure}	



For the purposes of finer assignment of resources, BSs collect with a regular interval of $\Delta t_s$, the scheduling requests (SRs) sent by the vehicles. SR contains the identifications (IDs) of the Tx- and the Rx-vehicles (in the case of one-to-one transmissions), their velocities, as well as $T_p$ of the V2V messages to be transmitted in DOCA. The collected information is then used by the controlling entity to predict future trajectories of the vehicles for the time which they will be inside DOCA. Predicted position information is used by the scheduler as we describe in the following. 



Regarding Case I, the scheduling task is trivial. Namely, for each requested transmission, the scheduler can only assign a new RB in order to avoid any collision with the other transmissions taking place inside the DOCA. Considering the Case II, reuse of the RBs is possible among different collision domains within the DOCA, which requires a decision mechanism assigning them in a reliable way. We elaborate on the latter, as follows.

For each incoming SR, the scheduler goes through the requested transmissions starting from the first arrived one, and attempts to assign each transmission to an RB that does not violate the constraints of reliability and half-duplex. 

Starting from the first among $F$ subchannels at the requested time to transmit, an RB is assigned if all of the following apply: i) the Rx-vehicle is within the broadcast range of the Tx-vehicle; ii) the Tx- and Rx-vehicles are not previously scheduled for any other reception or transmission (half-duplex constraint); and iii) no other vehicle scheduled for a transmission in that RB is 
closer than $d$ to the Rx-vehicle; and iv) the Tx-vehicle is not within $d$ of a vehicle that was previously scheduled for another reception in that RB. In case none of the $F$ subchannels are available at the requested time to transmit, the scheduler continues by checking the RBs in the next TTI, which results in a delay of one $\Delta t$ for the transmission.

%
%


An example schedule for a DOCA of single-collision domain is shown in Fig. \ref{sch}. Vehicles A, B and C send SRs to the BS respectively, requesting transmissions with different $T_p$, which are all assumed to collide if assigned to the same RB. Therefore, the second message of vehicle B, B\textsubscript{2}, is scheduled in the next available RB at f\textsubscript{1}. C\textsubscript{2}, requested for t\textsubscript{4}, could only be scheduled in the next TTI (t\textsubscript{5}, f\textsubscript{0}), hence it experiences a delay, since all subchannels are occupied at the requested TTI. 


We introduce another constraint: the maximum amount of tolerable delay $T_d$ of the system, which could be imposed by the V2X services. If a message has to be delayed for longer than a duration comparable to its $T_p$, then the message is dropped, i.e., not admitted to the schedule. Such a situation may happen when there is a high demand on the radio resources among the vehicles, e.g., due to a larger $\gamma$. \journal{number of vehicles per specific area or many requested transmissions within the considered amount of time. Accordingly, no RBs within $T_d$ for a later-coming request may be available. In this case the requested message is not admitted for transmission.}$T_d$ is chosen to be in the order of $T_p$, since the next messages are expected to be updated in terms of content (e.g., the subsequent CAM would supersede the previous one).


The scheduler informs the vehicles about the schedule by sending scheduling assignments (SAs) timely before they enter the DOCA. SA is an array of values, as in Fig. \ref{sch}, where vehicles look up the RBs to transmit/receive the messages during their traversal of DOCA. On the other hand, they inform the BSs about their exit from DOCA\journal{either explicitly or implicitly (e.g., CAM transmissions received by BSs)}, so that the scheduler can 
better adapt the schedule for future transmissions.


%% file: analysis.tex
\subsection{Resource reservation for ``ad hoc'' services in DOCA}

``Ad hoc'' services occur at an unexpected time inside DOCA due to, e.g., an emergency break or crash notification, which implies that they cannot be pre-scheduled before the vehicles enter the DOCA. Consequently, we propose to reserve a portion of the available radio resources for such services, in order to still reliably support such services.

Assume that a regular cell in the network has a capacity of $N$ number of resources. Then, DOCA of equivalent size would have the same $N$ for all the transmissions inside; as such, being considered as equivalent to a regular cell.
\journal{If $P$ resources are taken by the pre-scheduled services and $A$ resources are occupied by the ad hoc services currently active in the DOCA, then at any point in time, we have:
\begin{align}
P[Overload]=\begin{cases}
1, & \text{if $N<P+A$}.\\
0, & \text{otherwise}.
\end{cases}
\end{align}
} Let $R$ be the number of reserved RBs for the ad hoc services within DOCA. If $A$ resources are occupied by the ad hoc services currently active in the DOCA, then, as long as $R<N$, we can write
\begin{align}
P[Overload]=\begin{cases}
1, & \text{if $R<A$}.\\
0, & \text{otherwise}.
\end{cases}
\end{align}

If we assume that ad hoc service arrivals follow Poisson distribution \cite{poisson}, we have

\begin{align}
P[k \geq R] = 1 - \sum_{k=0}^{R-1}e^{-\lambda} \frac{\lambda^k}{k!},
\end{align}
where $\lambda$ is the arrival rate of the ad hoc services within a single, one-dimensional collision domain, given by
\begin{align}
\lambda = \gamma \times min(2d,l) \times \lambda_{AdHoc},
\label{eq:lambda}
\end{align}
where $\gamma$ is the vehicle density (number of vehicles per unit distance) within DOCA, and $\lambda_{AdHoc}$ is the probability a vehicle generates an ad hoc service per unit distance. Depending on the size of the DOCA, i.e., its length $l$, it may contain one or more collision domains defined by the broadcast range $d$. For this, we distinguish between two cases:

\subsubsection{Case I: DOCA is a single collision domain}
In this case, we have $l \leq 2d$, hence Eq. \ref{eq:lambda} becomes 
$ \lambda = \gamma \times l \times \lambda_{AdHoc}$. Within the DOCA, the transmissions will interfere with each other if they use the same RB.

\subsubsection{Case II: DOCA is not a single collision domain}
In this case $l>2d$, and $\lambda$ does not grow above $2d$, i.e., $ \lambda = \gamma \times 2d \times \lambda_{AdHoc}$. Instead, Poisson arrivals follow a memoryless property for each collision domain, and different services within DOCA can use the same RB, if they are taking place far enough from each other (i.e., at different collision domains). 
In other words, in Case II, the spatial reuse of radio resources is possible.


%

%

If we define the reliability of ad hoc services as $Rel = 1-P[Overload]$, we have
\begin{align}
Rel = e^{-\lambda}\sum_{k=0}^{R-1} \frac{\lambda^k}{k!}.
\label{eq:R}
\end{align}

If we solve Eq.~\ref{eq:R} for $R$, we determine the required amount of resources needed to achieve a given target reliability $Rel_{target}$ for ad hoc services.

After the above procedure, there remains $N-R$ resources to be used for services that can be pre-scheduled. Such services (e.g., platooning, CAM transmissions, etc.) include planned transmissions which are usually periodic, allowing us for a more granular allocation of the resources. The following subsection explains our algorithm to schedule these services within DOCA.

%% file: Results.tex
\section{Results and Evaluation}
\label{Results}

\subsection{Simulation Setup}

The proposed system as described in Sections \ref{Model} and \ref{Algorithm} is implemented in MATLAB. The implemented model consists of the sytem-level parameters summarized in Table \ref{table} with their default values. \journal{Unless otherwise mentioned, the default values are used.}

\begin{table}[!t]
	\renewcommand{\arraystretch}{1.1} 
	\caption{Simulation Parameters}
	\label{table}
	\resizebox{\columnwidth}{!}{%
		\centering
		\begin{tabular}{|l|l|}
			\hline
			Length of the DOCA, $l$ & 1000 m\\
			\hline
			Probability of ad hoc events, $\lambda_{AdHoc}$ & 0.05 events/vehicle/m\\
			\hline
			Broadcast/interference range, $d$ & 75 m\\
			\hline
			Arrival rate of vehicles at the DOCA, $\lambda_{arr}$ & 3 vehicles/s\\
			\hline
			Direction of an arriving vehicle & From either ends of the DOCA with equal probability\\ 
			\hline
			TTI duration, $\Delta t$ & 0.25 s\\
			\hline
			Number of subchannels, $F$ & 5\\
			\hline
			Message periodicity of each vehicle, $T_p$ & $T_{p}=k \Delta t$ s, $k=\{1,2,3\}$ with equal probability\\ 
			\hline
			Maximum allowed delay, $T_d$ & $4\Delta t=1$ s\\
			\hline
			SR collecting periodicity, $\Delta t_s$ & 0.3 s\\
			\hline	
		\end{tabular}
	}
\end{table}

\subsection{Resource reservation for ad hoc services}

\begin{figure}[!t]
	\centering
	\includegraphics[trim=1cm 20cm 1cm 20cm,clip=true,width=\columnwidth]{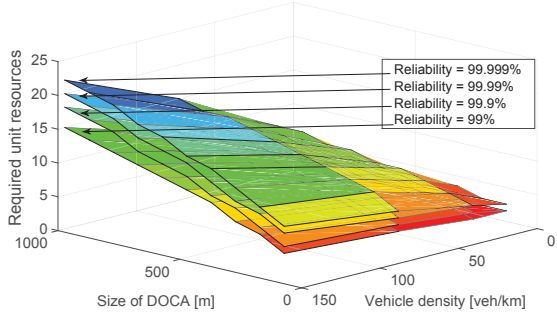}
	\caption{$R$ as a function of $l$ and $\gamma$, with respect to different $Rel_{target}$}
	\label{resultsR}
\end{figure}

Figure~\ref{resultsR} shows the result of numerical simulations for required $R$ to support ad hoc services under given reliability requirements. Specifically, we assume a perfect resource allocation: one that assigns the ad hoc services in non-overlapping resources without any scheduling overhead. In other words, the results in Fig.~\ref{resultsR} present the best case, with the minimum number of resources reserved for a target reliability. 

The results indicate that the increase in reliability does not penalize the system prohibitively. This is in contrast with the efficiency penalty on the physical layer, where increase in reliable transmissions would be costlier in terms 
of the spectral efficiency \cite{verdu}. 
Furthermore, Fig.~\ref{resultsR} shows that $\gamma$, as well as $l$, have a more significant effect on the required resources than the target reliability. The results provide design guidelines for a DOCA resource allocation, which should be sensitive to vehicle density changes and adapt both the amount of resources reserved as well as the schedule according to the density and vehicle mobility in DOCA.

\subsection{Impact of the predictions and vehicle velocities on the scheduling performance}
In this subsection,\journal{we evaluate the impact of different types of predictions regarding the vehicle velocities, on the scheduling performance.} rather than concentrating on how the predictions are made, we analyze the consequences of different types of predictions 
on the scheduling performance\journal{, where the vehicle velocities are predicted to be less than, equal to or over the actual values}. For this, we consider two cases: 1) all vehicles travel with a constant speed of 30 m/s; and 2) vehicles have random constant speeds uniformly distributed between 20 and 30 m/s.

For the first case, we evaluate the performance of the scheduler when the vehicles are predicted to have the same constant speed of 5, 15, 30, 35 and 45 m/s, as well as random constant speeds uniformly distributed between 5-15, 15-25, 25-35, 35-45 and 45-55 m/s. For the second case, we evaluate the effect of vehicles being predicted to have the same constant speed of 5, 15, 25, 35 and 45 m/s, together with the predictions of random constant speeds distributed uniformly between 5-15, 20-30, 25-35, 35-45 and 45-55 m/s.

\journal{From the predicted speeds, BSs calculate the positions of the vehicles during which they transmit/receive the V2V messages within the DOCA. Using the position information, scheduling by the BSs is performed according to the algorithm described in Section \ref{Scheduling}.} All pre-scheduled services are assumed to be periodic unicast (one-to-one) messages. In particular, each vehicle has a message traffic with random $T_{p}$, to be transmitted to the vehicle following behind it at the time it is entering the DOCA, and desires to maintain this communication for the rest of the time they are inside the DOCA together.

\begin{figure}[!t]
	\centering
	\includegraphics[width=\columnwidth]{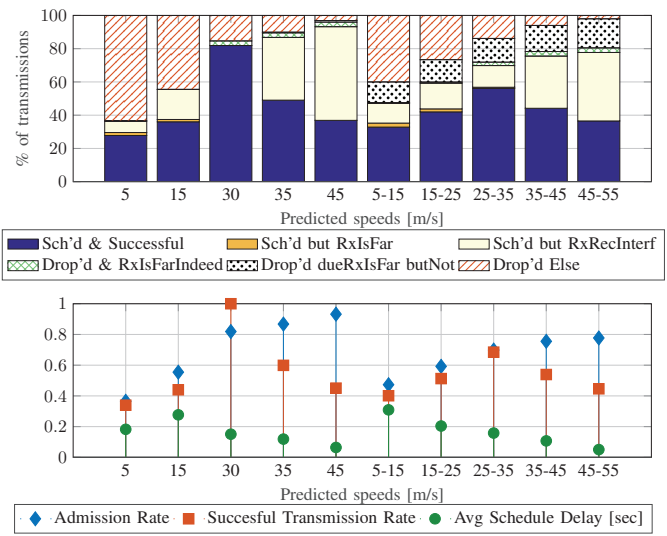}
	\caption{Impact of speed predictions on the scheduling performance. All vehicles have the same speed: 30 m/s.}
	\label{Same}
\end{figure}
\begin{figure}[!t]
	\centering
	\includegraphics[width=\columnwidth]{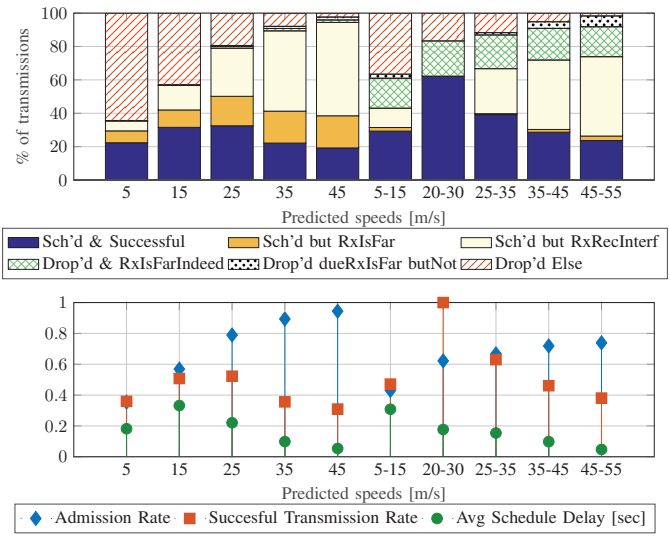}
	\caption{Impact of speed predictions on the scheduling performance. Vehicles have random constant speeds, uniformly distributed between 20 and 30 m/s.}
	\label{Different}
\end{figure}

The results are provided in Fig. \ref{Same} and \ref{Different}, respectively for the cases 1) and 2). Performance of the scheduler is measured via several key performance indicators (KPIs) with respect to the above cases of predicted and actual velocities of the vehicles. The determined KPIs are listed in Table \ref{KPIs}.

\begin{table}[!t]
	\renewcommand{\arraystretch}{1.1} 
	\caption{Scheduling KPIs}
	\label{KPIs}
	\resizebox{\columnwidth}{!}{%
		\centering
		\begin{tabular}{|l|}
			\hline
			\multicolumn{1}{|c|}{\textbf{Percentage of transmissions classified as:}}\\
			\hline
			\hline
			\textbf{Sch'd \& Successful:} scheduled, transmitted and successfully received\\
			\hline
			\textbf{Sch'd but RxIsFar:} scheduled and transmitted, however the Rx-vehicle is\\ 
			actually outside the transmission range, hence not successfully received \\
			\hline
			\textbf{Sch'd but RxRecInterf:} scheduled and transmitted, however the Rx-vehicle is\\ 
			actually subject to interference due to any other Tx-vehicle within $d$ away from it,\\
			hence not successfully received\\
			\hline	
			\textbf{Drop'd \& RxIsFarIndeed:} not admitted to the schedule since the Rx-vehicle is \\
			predicted to be outside $d$ at that instance, and this turns out to be true\\
			\hline
			\textbf{Drop'd dueRxIsFar butNot:} not admitted to the schedule due to the previous reason,\\
			however the Rx-vehicle is actually traveling within $d$ away from the Tx-vehicle\\
			\hline
			\textbf{Drop'd Else:} not admitted to the schedule due to any other reason, e.g., Rx-vehicle \\ is predicted to receive interference at that time instance\\
			\hline
			\hline
			\multicolumn{1}{|c|}{\textbf{Several other indicators:}}\\
			\hline
			\hline
			\textbf{Admission Rate:} the ratio of the number of scheduled transmissions to the \\
			total number of requested transmissions\\
			\hline
			\textbf{Successful Transmission Rate:} the ratio of the number of successful transmissions \\ that were requested, to the number of transmissions admitted in the case of \\
			a correct predictor (correctly predicting the actual velocities of the vehicles)\\
			\hline
			\textbf{Avg Schedule Delay:} the mean value of the delay experienced among all scheduled\\ transmissions, in seconds\\
			\hline
		\end{tabular}
	}
\end{table}

%
%
%

As expected and can be seen from Fig. \ref{Same} and \ref{Different}, correct predictions achieve the largest \textit{Sch'd \& Successful}, hence \textit{Successful Transmission Rate}. 
Accordingly, both KPIs decrease with the predicted speeds deviating from the actual values. To illustrate, when vehicles are predicted to be all traveling at 35 m/s instead of their actual speeds of 30 m/s, \textit{Successful Transmission Rate} decreases around 40\%.

Note that even with correct predictions not all transmissions could be scheduled (i.e., \textit{Admission Rate} is less than 1). This is because for some Rx-vehicles, it is not possible to schedule them given the system constraints $F$ and $T_d$, without any interference during at least some part of their time within DOCA, or they might not be within $d$ from the Tx-vehicle. It can be observed for the correct predictions in Fig. \ref{Different} that both occasions rise, as the relative speeds of the vehicles increased. 

Considering the cases where vehicles are predicted to be slower, the percentage of \textit{Drop'd Else} considerably increases, besides the transmissions \textit{Sch'd but RxRecInterf}, all due to the errors in the predicted positions of the interferers. Consequently, \textit{Admission Rate} can drop below 0.5 if the velocities are predicted as low as 5 m/s.

On the other hand, when the vehicles are predicted to be faster, \textit{Sch'd but RxRecInterf} are present with larger percentages, in addition to the occurrences of \textit{Drop'd \& RxIsFarIndeed} and \textit{Drop'd dueRxIsFar butNot}. This can be explained by our assumption that 
each vehicle transmits to the vehicle following itself. If the vehicle entered the DOCA is predicted to be faster, then the corresponding Rx-vehicle is thought as being left far behind it, hence the messages are (erroneously) dropped. Similarly, interferers are also thought to be away from the Rx-vehicles, resulting in higher \textit{Admission Rates}.

For the cases of vehicles having different relative speeds, as provided in Fig. \ref{Different},
the percentage of \textit{Drop'd \& RxIsFarIndeed} is more pronounced than \textit{Drop'd dueRxIsFar butNot}, due to Rx-vehicles now being able to 
overtake their Tx-vehicles, and even moving farther than $d$ apart. 
This also results in considerable percentage of \textit{Sch'd but RxIsFar}, especially if the vehicles are all predicted as having the same speed.

Regarding \textit{Avg Schedule Delay}, it is interesting to observe the trend where it decreases by predicting the vehicles to be faster. Such predictions assume less collisions, resulting in more admissions to the schedule, hence the transmissions experience less delay (although they eventually collide).

\journal{

\subsection{Impact of the Interference Radius}
In this subsection, we evaluate the impact of the size $d$ of the ``broadcast'' or the ``interference'' range on the schedule. The range is practically determined according to many factors in the network, such as the transmit power of the vehicle antenna, propagation losses and fading on the radio channel. Thus, it is possible for the system to operate under many different $d$. Here, we consider the cases of $d$ being equal to 45 m (results in Figures \ref{SameR45} and \ref{DifferentR45}) and 100 m (Figures \ref{SameR100} and \ref{DifferentR100}), in addition to the case of 75 m in the previous section (Figures \ref{Same} and \ref{Different}), and compare the measured KPIs of the schedules based on them.

As also the name suggests, ``interference range'' determines the size of the area where any vehicle contained within will receive interference from the Tx-vehicle, if it is not the intended receiver but using the same RB for its own V2V-message reception. Accordingly, increasing this range increases the number of vehicles receiving interference, given the same vehicle density on the road. The ratios of the scheduled transmissions for which the receiver gets interference or the ones dropped due to predicted interference, i.e., \textit{Sch'd but RxRecInterf} and \textit{Drop'd Else}, increase going through the Figures \ref{SameR45} and \ref{DifferentR45}, Figures \ref{Same} and \ref{Different}, and Figures \ref{SameR100} and \ref{DifferentR100}, respectively.

On the other hand, as our scenario assumes the Rx-vehicle to be the following vehicle behind the transmitter, the ratio of the transmissions that the intended receivers can not receive as they are more than $d$ away from the Tx-vehicle, i.e., the transmissions \textit{Sch'd but RxIsFar} and \textit{Drop'd \& RxIsFarIndeed} decrease, as a larger $d$ increases the possibility of ``covering'' the intended Rx-vehicle. Similarly, the Rx-vehicles are less likely predicted to be out of the ``broadcast range'' $d$, as the ratio of the transmissions \textit{Drop'd dueRxIsFar butNot} also decrease with the increased $d$.

\begin{figure}[!t]
	\centering
	\resizebox{\columnwidth}{!}{\input{images/D1000L3R45_sameSpeed.tikz}}
	\caption{Impact of different predictions on the scheduling performance, in terms of the measured KPIs, in the case of all vehicles having the same speed of 30 m/s with the size of interference range $d=45$.}
	\label{SameR45}
\end{figure}

\begin{figure}[!t]
	\centering
	\resizebox{\columnwidth}{!}{\input{images/D1000L3R45_differentSpeed.tikz}}
	\caption{Impact of different predictions on the scheduling performance, in terms of the measured KPIs, in the case of vehicles having different speeds uniformly-random distributed between 20 and 30 m/s with the size of interference range $d=45$.}
	\label{DifferentR45}
\end{figure}

\begin{figure}[!t]
	\centering
	\resizebox{\columnwidth}{!}{\input{images/D1000L3R100_sameSpeed.tikz}}
	\caption{Impact of different predictions on the scheduling performance, in terms of the measured KPIs, in the case of all vehicles having the same speed of 30 m/s with the size of interference range $d=100$.}
	\label{SameR100}
\end{figure}

\begin{figure}[!t]
	\centering
	\resizebox{\columnwidth}{!}{\input{images/D1000L3R100_differentSpeed.tikz}}
	\caption{Impact of different predictions on the scheduling performance, in terms of the measured KPIs, in the case of vehicles having different speeds uniformly-random distributed between 20 and 30 m/s with the size of interference range $d=100$.}
	\label{DifferentR100}
\end{figure}

\subsection{Impact of the Vehicle Density}
In this subsection, we evaluate the impact of the rate $\lambda_{arr}$, the number of vehicles arriving in both directions per second, on the schedule. From this rate, it is possible to determine the vehicle density $\gamma$ on the road, namely the two-way traffic volume in terms of the number of vehicles per a unit section of the highway, given a constant flow. Traffic volume is an important parameter in the design of transportation systems, regarding, e.g., the capacity of the roads. Even for the same road, it takes different values on an hourly or a seasonal basis. 

In our case, we are concerned about the radio resource usage by the vehicles on a given highway segment, which is mainly based on the requested number of transmissions. In fact, demand on the radio resources is proportionally related to $\lambda_{arr}$. Correspondingly, for a fixed value of $\lambda_{arr}$, increasing the frequency of V2V message traffic would also create the same effect on the system.

We are interested in how the schedule gets affected by different vehicle densities. For this, KPIs are provided for the values of $\lambda_{arr}=1$ and 5, respectively in Figures \ref{SameL1} and \ref{DifferentL1}, and \ref{SameL5} and \ref{DifferentL5}, besides the results for $\lambda_{arr}=3$ in Figures \ref{Same} and \ref{Different}. Other system-level parameters are kept constant, i.e., $l=1000$ m and $d=75$ m.

It is clear that, with the increased density of vehicles, the ratio of the transmissions not admitted to the schedule mostly due to the predicted interference, as well as the ones scheduled but the receiver getting interference, i.e., the transmissions \textit{Drop'd Else} and \textit{Sch'd but RxRecInterf}, is increased. Similar to the case of increasing $d$, the number of vehicles to suffer from interference due to any transmission taking place is essentially increased, but this time due to higher number of vehicles within the same transmission range. 

On the other hand, due to our system model, having a low rate of arriving vehicles may result in some Rx-vehicles being out of the broadcast range of the Tx-vehicles, by the time they arrive at the DOCA. This results in increased ratio of the transmissions \textit{Sch'd but RxIsFar}, \textit{Drop'd \& RxIsFarIndeed}, and \textit{Drop'd dueRxIsFar butNot}, similar to the effect of decreasing $d$. Such occurrences can be observed for the case of $\lambda_{arr}=1$, which can be seen in Figures \ref{SameL1} and \ref{DifferentL1}.

\begin{figure}[!t]
	\centering
	\resizebox{\columnwidth}{!}{\input{images/D1000L1R75_sameSpeed.tikz}}
	\caption{Impact of different predictions on the scheduling performance, in terms of the measured KPIs, in the case of all vehicles having the same speed of 30 m/s with the rate of arriving vehicles $\lambda_{arr}=1$ vehicle/second.}
	\label{SameL1}
\end{figure}

\begin{figure}[!t]
	\centering
	\resizebox{\columnwidth}{!}{\input{images/D1000L1R75_differentSpeed.tikz}}
	\caption{Impact of different predictions on the scheduling performance, in terms of the measured KPIs, in the case of vehicles having different speeds uniformly-random distributed between 20 and 30 m/s with the rate of arriving vehicles $\lambda_{arr}=1$ vehicle/second..}
	\label{DifferentL1}
\end{figure}

\begin{figure}[!t]
	\centering
	\resizebox{\columnwidth}{!}{\input{images/D1000L5R75_sameSpeed.tikz}}
	\caption{Impact of different predictions on the scheduling performance, in terms of the measured KPIs, in the case of all vehicles having the same speed of 30 m/s with the rate of arriving vehicles $\lambda_{arr}=5$ vehicles/second.}
	\label{SameL5}
\end{figure}

\begin{figure}[!t]
	\centering
	\resizebox{\columnwidth}{!}{\input{images/D1000L5R75_differentSpeed.tikz}}
	\caption{Impact of different predictions on the scheduling performance, in terms of the measured KPIs, in the case of vehicles having different speeds uniformly-random distributed between 20 and 30 m/s with the rate of arriving vehicles $\lambda_{arr}=5$ vehicles/second..}
	\label{DifferentL5}
\end{figure}

\subsection{Impact of the DOCA Size}
The size of the DOCA determines the duration for which the vehicles spend inside it. In our preliminary system model, the scheduler calculates the positions of the vehicles based on their predicted velocities, which is assumed to be constant over time. Accordingly, the deviation between the actual and the predicted positions of the vehicles increases with time, in the case of inaccurately predicted velocities. 

Increasing the size of the DOCA, which is the length $l$ in our scenario, results in higher ratios of the transmissions that are either not received or not scheduled due to wrong predictions on the positions of the Rx-vehicles, i.e., the transmissions \textit{Sch'd but RxIsFar} and \textit{Drop'd dueRxIsFar butNot}.

} 

%% file: Conclusions.tex
\section{Conclusions and Future Work}
\label{Conclusion}

In this work, we considered a special use case of V2V communications, where vehicles communicate inside DOCA, an out-of-coverage area that is delimited by network infrastructure. For this, a system is proposed in which BSs delimiting the area reserve and schedule radio resources for the vehicles. In order to satisfy reliability constraints required by V2V services, the controller is proposed to make predictions (e.g., regarding the vehicle positions). We analyzed the resource reservation and performance of the scheduler in terms of reliability, admission rate, and delay. 

Impact of other system-level parameters such as $l, d, \lambda_{arr}$ on the scheduling performance is not provided due to space limitations. Moving forward, more diversity could be introduced into the system in terms of message and road traffic, e.g., including broadcast/multicast transmissions, and 
more varied propagation conditions. 
Second, predictions that in this work are assumed as a given, could be instead made by a scheduling algorithm, 
where past trajectory of the vehicles, as well as environmental parameters such as vehicle type, time of the day, etc. are exploited to make better scheduling predictions.